\documentstyle[aasms4]{article}
\begin{document}
\title{A Robust Age Indicator for Old Stellar Populations}
\author{A. Vazdekis \& N. Arimoto}
\affil{Institute of Astronomy, School of Science,
University of Tokyo, Osawa 2-21-1, Mitaka, Tokyo 181-8588, Japan\\
(e-mail: vazdekis@mtk.ioa.s.u-tokyo.ac.jp, arimoto@mtk.ioa.s.u-tokyo.ac.jp)}
\journalid{Vol}{Journ. Date}
\articleid{start page}{end page}
\paperid{manuscript id}
\cpright{type}{year}
\ccc{code}
\lefthead{Vazdekis \& Arimoto}
\righthead{A robust age indicator for old stellar populations}

\begin{abstract}
We derive new spectral H$\gamma$ index definitions which are robust age 
indicators for old and relatively old stellar populations and therefore have 
great potential for solving the age-metallicity degeneracy of galaxy spectra. 
To study this feature as a function of age, metallicity and resolution, we have 
used a new spectral synthesis model which predicts spectral energy distributions
of single-age, single-metallicity stellar populations at resolution 
FWHM$\sim1.8$\AA~(which can be smoothed to different resolutions), allowing
direct measurements of the equivalent widths of particular absorption features. 
We show that H$\gamma$ strong age disentangling power is due to a compensating 
effect: at specified age, H$\gamma$ strengthens with metallicity 
due to an adjacent metallic absorption, but on the other hand the adopted 
pseudocontinua are depressed by the effects of strong neighboring Fe{\sc I} 
lines on both sides of H$\gamma$. Despite the fact that this effect depends 
strongly on the adopted resolution and galaxy velocity dispersion $\sigma$, 
we propose a system of indicators which are completely insensitive to 
metallicity and stable against resolution, allowing the study of galaxies up 
to $\sigma\sim300$~kms$^{-1}$.

An extensive analysis of the characteristics of these indices indicates that 
observational spectra of very high signal-to-noise ratio and relatively high
dispersion, are required to gain this unprecedented age discriminating power. 
Once such spectra are obtained, accurate and reliable estimates for the 
luminosity-weighted average stellar ages of these galaxies will become possible 
for the first time, without assessing their metallicities. We measured this
index for two globular clusters, a number of low-luminosity elliptical galaxies 
and a standard S0 galaxy. We find a large spread in the average stellar ages of 
a sample of low-luminosity ellipticals. In particular, these indices yield
4~Gyr for M~32. This value is in excellent agreement with the age provided by
an extraordinary fit to the full spectrum of this galaxy that we achieve 
in this paper.

\end{abstract}


\section {Introduction}
The understanding of the stellar populations of early-type galaxies plays a key 
role in assessing the origin of these systems. Elliptical galaxies yield robust 
color-magnitude relations, as shown with high photometric precision by Bower, 
Lucey \& Ellis (1992) for the Virgo and Coma clusters. The origin of this
color-magnitude relation is under a strong debate; the relation could either be 
caused by a variation of the mean stellar metallicity along the relation 
(e.g., Arimoto \& Yoshii 1987; Kodama \& Arimoto 1997), or it can be attributed
to a variation of the age and metallicity (Ferreras, Charlot \& Silk 1998). This
is because the interpretation of the stellar populations from the integrated 
light of galaxies suffers from the fact that there is an age-metallicity 
degeneracy, i.e, the two effects cannot be separated simultaneously with 
current techniques (O'Connell 1986; Worthey 1994; Arimoto 1996). 

One expects that, rather than using colors, more accurate spectral information 
should be able to break this degeneracy. In particular, Balmer lines were 
thought to be ideal candidates, because younger (i.e. hotter) stars show 
stronger hydrogen absorption (O'Connell 1976). However, Worthey (1994) showed 
that use of H$\beta$ at intermediate spectral resolution 
($\sigma\sim200$~kms$^{-1}$) together with a large set of the other Lick 
metallic absorption lines is not enough to provide the desired age-metallicity 
discrimination. Therefore more accurate and reliable estimates for the ages of 
early-type galaxies are crucial in understanding their formation. 
Worthey \& Ottaviani (1997) showed the importance of increasing resolution when 
working with Balmer lines, and Jones \& Worthey (1995) (hereafter JW95) have 
shown the ability of their H$\gamma_{HR}$ (a redefinition of Rose 1994 H$\gamma$
index) high resolution index to separate the metallicity and age effects.

In this paper we study H$\gamma$ feature to understand its power in breaking the
well known degeneracy. In \S~2 we use the high spectral resolution 
(FWHM$\sim$1.8\AA) single-age, single-metallicity stellar populations (SSP's) 
model library of Vazdekis (1999) (hereafter V99) and find that H$\gamma_{HR}$
is strongly dependent on resolution and galaxy velocity dispersion. 
In \S~3 we study H$\gamma$ as a function of age, metallicity and resolution and 
define a new indicator which is completely insensitive to metallicity and stable
against resolution. We deal with the most important problems affecting this 
indicator and propose a recipe for determining the average stellar age of a 
galaxy. In \S~4 we present a discussion and in \S~5 our conclusions.

\section {The H$\gamma$ feature}

\subsection {The SSP spectral library}

To study H$\gamma$ feature we use the spectral synthesis model of V99 (an 
extended version of the evolutionary stellar population synthesis model of 
Vazdekis et al. 1996) which predicts spectral energy distributions (SED's) 
for single-aged old stellar populations with metallicities 
$-0.7\leq[Fe/H]\leq+0.2$ at resolution FWHM$\sim1.8$\AA~in the optical region. 
The model uses as input database the empirical spectral library of Jones (1997),
after a careful selection of a subsample of $\sim$550 stars. 
One important advantage of this model is that it allows direct measurements of 
particular absorption features on the spectra of SSP's with specified 
metallicity, age, and initial stellar mass function (IMF), instead of employing 
polynomial fitting functions that relate the absorption line-strengths with the 
stellar atmospheric parameters (Worthey 1994; Vazdekis et al. 1996). We use the 
{\it bimodal IMF} of Vazdekis et al. (1996) with a shallow low-mass 
($\leq 0.6 M_{\odot}$) slope and a Salpeter (1955) high mass slope.

\subsection {The behavior of H$\gamma$ as a function of resolution}
In V99 we showed the intermediate resolution indices H$\gamma_{A}$ and 
H$\gamma_{F}$ (Worthey \& Ottaviani 1997) and the high resolution index 
H$\gamma_{HR}$ (JW95) as a function of the age and metallicity for SSP's. 
Unexpectedly, the two set of indices show opposite trends: at specified age, 
most metal-deficient SSP's give largest values for H$\gamma_{A}$ and H$\gamma_{F}$ 
but H$\gamma_{HR}$. Moreover, the wavelength baselines covered by H$\gamma_{A}$ 
and H$\gamma_{F}$ are considerably larger than the one by H$\gamma_{HR}$. This 
fact suggests that H$\gamma$ is not only sensitive to resolution but also to 
the way in which the index is defined, as was pointed out by JW95. We therefore 
speculate that there must be a particular definition (with wavelength coverage
larger than in H$\gamma_{HR}$ but shorter than in H$\gamma_{F}$ and 
H$\gamma_{A}$, where the index is insensitive to metallicity. Fig.~1 shows 
H$\gamma_{HR}$ as a function of the metallicity and age of the SSP for 
different resolutions $\sigma$'s. For $\sigma=60$~kms$^{-1}$ the most metal-rich
SSP's provide the largest values. This tendency is much less pronounced 
for $\sigma=125$~kms$^{-1}$, where the previous trend is starting to change 
(e.g., the values for [Fe/H]=$-$0.4 are larger than those for [Fe/H]=0.0). For 
$\sigma=200$~kms$^{-1}$ we get the maximum convergence of the lines; this 
resolution seems to be the {\it inflection point} at which all loci with 
different metallicities converge. Finally, for $\sigma=275$~kms$^{-1}$ the 
trend is completely inverted (resembling H$\gamma_{A}$ and H$\gamma_{F}$).

Fig.~2 shows various SSP model spectra of 13~Gyr and different metallicities 
broadened to the resolutions used in Fig.~1. In first panel we mark the definition
of H$\gamma_{HR}$, which measures the EW of a total area of 3.74~\AA~centered on 
H$\gamma$ by selecting as pseudocontinua the left and right peaks (indicated by 
arrows). While H$\gamma$ has nearly the same depth, the metallic contribution
from its left side is strengthening the feature for more metal-rich populations. 
For stars, the dependence of H$\gamma_{A}$, H$\gamma_{F}$, and H$\gamma_{HR}$ as
a function of T$_{eff}$ was shown by Worthey \& Ottaviani (1997) and Jones (1997)
respectively: in all cases the EW increases very sharply for T$_{eff}$'s larger 
than $\sim$5500~K, but for lower T$_{eff}$'s where H$\gamma$ is basically 
constant. Since old stellar populations have T$_{eff}$'s in this lower
temperature range, differences in T$_{eff}$ due to changes in metallicity are 
not enough to vary the absorption strength rapidly enough to compensate the very 
slow change of H$\gamma$. We conclude that for old stellar populations the 
strengthening of H$\gamma_{HR}$ as a function of metallicity is mainly caused 
by its adjacent blue side metallic absorption. When broadening the spectra to 
$\sigma=125$~kms$^{-1}$ these metallic features are merged with H$\gamma$ line. 
However the adopted pseudocontinua start to be depressed by the deepening of 
the strong neighboring Fe{\sc I} lines centered on 4325~\AA~and 4352~\AA~(the 
latter also contains Mg{\sc I}), affecting H$\gamma_{HR}$ strength in the 
reverse sense. For $\sigma=200$~kms$^{-1}$ these pseudocontinua are more 
depressed (than for $\sigma=125$~kms$^{-1}$) for more metal-rich SSP's, making 
H$\gamma_{HR}$ nearly identical. This is what we have called the inflection 
point, where metallicity effects are negligible. Finally, for 
$\sigma=275$~kms$^{-1}$ the pseudocontinua are considerably depressed, 
breaking the effect that occurs at $200$~kms$^{-1}$, and causing larger EW's 
for most metal-deficient SSP's. Unfortunately, due to limitations in model 
parameter coverage, we are unable to study in detail the inflection point for 
SSP's of younger ages and/or lower metallicities.

\section {The new age indicator}
\subsection {New index definitions}
Fig.~1 shows that H$\gamma_{HR}$ provides a stronger power at separating 
age at $\sigma=200$~kms$^{-1}$ than for lower or higher resolutions. JW95 
calculated  H$\gamma_{HR}$ for SSP's using a different approach: they derived 
empirical fitting functions which relate the stellar atmospheric parameters to 
the measured line-strengths on the basis of Jones (1997) stellar library (after
broadening the spectra to $\sigma=83$~kms$^{-1}$). Their H$\gamma_{HR}$ model 
predictions fall among the first two plots of Fig.~1 (i.e. those calculated for 
$\sigma = 60$ and $125$~kms$^{-1}$). Fig.~1 also shows that H$\gamma_{HR}$ is 
very sensitive to resolution due to the selection of these two peaks as 
pseudocontinua (which also vary in height and position as a function of 
metallicity). Thus, H$\gamma_{HR}$ age disentangling power is destroyed with 
very small $\sigma$ changes.

To achieve a metallicity insensitive H$\gamma$ indicator which is stable 
against $\sigma$ we used the following approach: first we broadened the SSP 
spectral library of V99 by steps of $25$~kms$^{-1}$ from $\sigma=60$~kms$^{-1}$ to 
$\sigma=400$~kms$^{-1}$. We propose various H$\gamma$ index definitions,
and then we measure them on these model spectra. We find new indices, 
[H$\gamma$+1/2(Fe{\sc I}+Mg{\sc I})]$_{\sigma}$, which are completely 
insensitive to metallicity, at well separated resolutions: 
$\sigma = 125$~kms$^{-1}$, $200$~kms$^{-1}$ and $275$~kms$^{-1}$, each quite 
stable in a range of $\Delta\sigma\sim75$~kms$^{-1}$. Table~1 provides the 
new definitions (also marked in Fig.~2).

[H$\gamma$+1/2(Fe{\sc I}+Mg{\sc I})]$_{125}$ uses the effect of the blue 
pseudocontinuum to precisely guarantee this ability to disentangle the age, 
while for the red pseudocontinuum it uses a portion of the very stable 
continuum around 4364\AA. To achieve the resolution stability (required for 
practical applications) we select the blue pseudocontinuum to fall into the 
bottom of H$\gamma$ ($\lambda$=4340.468\AA), and the feature to fall well 
inside this pseudocontinuum. This selection introduces a change that is 
cancelled by extending the red edge of our feature to the bottom of 
Fe{\sc I}+Mg{\sc I} at $\lambda$=4352.737\AA~(Fe{\sc I}): for increasing 
metallicities the strength lost due to blue pseudocontinuum depression
(caused by the strengthening of Fe{\sc I} at 4325\AA~and by the full inclusion 
in this pseudocontinuum of H$\gamma$ blue side metallic absorption) is 
cancelled with the depth gained by Fe{\sc I}+Mg{\sc I} strengthening (which is 
contributing to the feature in the new definition). Thus this index is not 
just an EW as we are dealing with the {\it absorption line profile}, and then 
it is required spectroscopy of very high signal-to-noise ratio and 
relatively high dispersion (see \S~3.2). At this level of accuracy, an 
influence of the stellar kinematics cannot be neglected. In particular, 
kinematically complex systems (e.g., Prada et al. 1996) would require a 
simultaneous line profiles and kinematics processing. 

Modifying these definitions slightly, we find similar age-metallicity trends for
$\sigma=200$~kms$^{-1}$ and $\sigma=275$~kms$^{-1}$, providing equally stable 
$\sigma$ ranges. These indices allow us to cover almost the whole range 
of galaxy $\sigma$'s up to $\sim300$~kms$^{-1}$. However, despite the fact that 
such inflection point can essentially be found for larger $\sigma$'s 
($\sim400$~kms$^{-1}$), resulting solutions were not stable against resolution.
Fig.~3 demonstrates the robustness and age disentangling power of
[H$\gamma$+1/2(Fe{\sc I}+Mg{\sc I})]$_{\sigma}$ system. Bottom-left panel of
Fig.~3 shows [H$\gamma$+1/2(Fe{\sc I}+Mg{\sc I})]$_{125}$ stability against 
resolution. Table~2 tabulates the mean index values for various ages of the 
SSP's. We also note that [H$\gamma$+1/2(Fe{\sc I}+Mg{\sc I})]$_{\sigma}$ indices
are probably not unique for achieving this age discriminating power and 
alternative definitions could well be sought.

\subsection {Index characteristics and major uncertainties}
In this section we study the problems that can affect 
[H$\gamma$+1/2(Fe{\sc I}+Mg{\sc I})]$_{\sigma}$ indicators for practical 
applications. Table~3 summarizes their characteristics and major uncertainties.

\subsubsection {Spectral resolution}
The spectral resolution effect is extensively discussed in \S~3.1. Table~3 
tabulates $\sigma$ ranges where [H$\gamma$+1/2(Fe{\sc I}+Mg{\sc I})]$_{\sigma}$
indices are stable against resolution. Within these ranges 
($\Delta\sigma\sim45$~kms$^{-1}$) it is possible to distinguish models of 13~Gyr 
and 17~Gyr, irrespective of the metallicity (in the range 
$-$0.7$\leq$[Fe/H]$\leq$+0.2). Bottom-left panel of Fig.~3 shows that                     
[H$\gamma$+1/2(Fe{\sc I}+Mg{\sc I})]$_{\sigma}$ indices are also stable within 
$\Delta\sigma\sim75$~kms$^{-1}$ but the age disentangling power decreases 
slightly (only SSP's younger than 12~Gyr can be distinguished from one of 17~Gyr). 

\subsubsection {Signal-to-noise ratio}
Unfortunately, the required signal-to-noise ratio per \AA, S/N(\AA), is very 
high, limiting the number of galaxies observable with most of present-day 
intermediate class telescopes. An estimate of the required exposure time can be 
done using the analytical approach of Cardiel et al. (1998) (eqs. (9) and (41) 
to (44)). Following their notation, coefficients c1 and c2 are calculated from 
eqs.~(43) and (44) using Table~1 index definitions. Assuming a desirable 
index error ($\sigma[I_{a}]$), we use eq. (41) to estimate the required 
S/N(\AA) (SN). Then by approaching 
$\sigma_{i}(\lambda_{i})\sim\sqrt{S(\lambda_{i})}$, we use eq. (42) to obtain 
the number of counts per ADU:
\begin{equation}
c=\frac{SN\theta}{2g}\left[SN+\sqrt{SN^{2}+\frac{4\sigma_{RN}^{2}}
{\theta}}\right] ,
\end{equation}
where $\theta$ is the dispersion in \AA~per resolution element, $g$ is the gain 
in $e^{-1}/ADU$, $\sigma_{RN}$ is the readout noise in $e^{-1}$. Finally we 
compare with the signal obtained in real observational runs. An example in 
which we integrate all photons within $\sim2\arcsec$ for a nearby galaxy center 
with $\mu_{B}=17.5~mag~arcsec^{-2}$ on a 4m class telescope is given in Table~3. 
We used $\theta=$0.8~\AA/pix and $\sigma_{RN}=4.4~e^{-1}$. Table~3 also shows 
that higher resolution [H$\gamma$+1/2(Fe{\sc I}+Mg{\sc I})]$_{\sigma}$ 
definitions require lower S/N's.

\subsubsection {Wavelength and radial velocity uncertainties}
Table~3 shows largest $\lambda$ errors or shifts allowed to guarantee 
[H$\gamma$+1/2(Fe{\sc I}+Mg{\sc I})]$_{\sigma}$ age disentangling power. Thus
irrespective of galaxy $\sigma$'s, observational spectra of relatively high 
dispersion ($\stackrel{<}{_\sim}$0.8\AA) are required to achieve an accurate 
$\lambda$ calibration that is typically $\sim$ 5-10\% of the adopted dispersion. 
This also implies that radial velocity effects such as galaxy internal 
rotational velocity or redshift should be taken into account with high 
precision. For this purpose the observed spectrum should be 
crosscorrelated with an appropriate V99 model spectrum (see Appendix). We also
note that a galaxy with a recession velocity of $\sim1000$~kms$^{-1}$ produces a
difference of $\sim0.13$\AA~between the bluest point of the blue pseudocontinuum
and the reddest point of the red pseudocontinuum, i.e., as large as the largest 
$\lambda$ shift allowed for [H$\gamma$+1/2(Fe{\sc I}+Mg{\sc I})]$_{125}$. 
Therefore the (1+$z$) effect should be properly accounted for.

\subsubsection {The spectrum shape} 
The effect of the spectrum shape on 
[H$\gamma$+1/2(Fe{\sc I}+Mg{\sc I})]$_{\sigma}$
should be small because the wavelength range involved is quite narrow. We
have done the following test: we removed the continuum of V99 model spectra 
(using an spline3 of order 6) and then measured 
[H$\gamma$+1/2(Fe{\sc I}+Mg{\sc I})]$_{\sigma}$. Table~3 shows the largest 
differences (obtained for the oldest SSP's) when comparing to previous 
measurements (i.e., on the SSP spectra with a flux calibrated response). 
We conclude that the effect of the instrumental response curve on 
[H$\gamma$+1/2(Fe{\sc I}+Mg{\sc I})]$_{\sigma}$ is nearly negligible. 

\subsubsection {Age uncertainties} 
Table~3 summarizes the largest theoretical age uncertainties as well as the 
ones obtained when adopting the largest errors allowed either in $\sigma$, 
$\Delta\lambda$ (e.g., errors in $\lambda$ calibration, redshift, rotation curve) 
or S/N slightly lower than recommended. 

\subsection {A recipe for determining the luminosity-weighted average stellar age of 
a galaxy}
A detailed description of the main characteristics and uncertainties of 
[H$\gamma$+1/2(Fe{\sc I}+Mg{\sc I})]$_{\sigma}$ indices was performed in \S~3.2.
Here we summarize the steps required to ensure a correct index measurement: 

\begin{enumerate}

\item The observed spectrum should satisfy the high S/N requirement 
of Table~3, which can be achieved by integrating over the spatial direction. 
Dispersion should be large enough (preferably $<$0.8\AA/pix) to prevent 
absolute $\Delta\lambda$ shifts or errors with a precision as given in Table~3. 

\item Determine $\sigma$ along the spatial direction (e.g., by using stellar 
template spectra acquired with the same instrumental configuration, or by using 
V99 models). Very accurate $\sigma$ measurement is not essential.

\item Crosscorrelate the observed spectrum with an appropriate V99 model
spectrum (instead of using a stellar template) to determine its absolute radial
velocity, which has to be taken into account when measuring the index (see 
Appendix). This is a very important step.

\item Integrate over spatial direction to achieve the required S/N. Variations 
of velocity rotation (and $\sigma$) must be smaller than indicated in Table~3.
Otherwise they should be carefully taken into account. 

\item Measure [H$\gamma$+1/2(Fe{\sc I}+Mg{\sc I})]$_{\sigma}$ index appropriate 
for obtained $\sigma$ and compare with Table~2 to estimate the 
age\footnote{If obtained $\sigma$ falls outside 
[H$\gamma$+1/2(Fe{\sc I}+Mg{\sc I})]$_{\sigma}$ resolution stability range (see 
Table~3), results improve slightly if the observational spectrum is broadened 
(by convolving with a gaussian) to match $\sigma$ of the index. However the 
reader should be aware that higher resolution definitions require lower 
S/N's (see Table~3).}. A fortran code is provided at our web site: 
http://www.ioa.s.u-tokyo.ac.jp/$\sim$vazdekis/. 

\end{enumerate}

\section{Discussion}

First panel of Fig.~3 shows [H$\gamma$+1/2(Fe{\sc I}+Mg{\sc I})]$_{125}$ 
measurements for two metal-rich globular clusters of our Galaxy, 47~Tuc and 
NGC~6624 (Rose 1994), five low velocity dispersion ellipticals of Jones (1997) 
(NGC~4489, NGC~4239, NGC~3605, NGC~4387 and M~32) and the standard S0 galaxy 
NGC~7332 (Vazdekis 1996). The latter spectrum, with 1~$hour$ exposure time, 
was obtained at Observatorio del Roque de Los Muchachos, La Palma, using the 
4.2m WHT (ISIS spectrograph). We integrate the innermost 1.5$\arcsec$ along 
the slit (positioned on the minor axis of the galaxy) to achieve a relatively 
high (although not enough) S/N(\AA)$\sim$140. Rotation curve and $\sigma$ were 
both found to be very stable in this region. All these spectra (except 
NGC~7332 which has $\sigma\sim135$~kms$^{-1}$) were pre-broadened to a common 
resolution of $125$~kms$^{-1}$. 

The M~32 spectrum (with extremely high S/N, see JW95) was broadened to 
$\sigma=200~kms^{-1}$ and $275~kms^{-1}$ for measuring 
[H$\gamma$+1/2(Fe{\sc I}+Mg{\sc I})]$_{200}$ and 
[H$\gamma$+1/2(Fe{\sc I}+Mg{\sc I})]$_{275}$ respectively (see right panels 
of Fig.~3). Obtained ages (using Table~2) are in good internal consistency 
since the three indices provide $\sim$4~Gyr. JW95 obtained $\sim$7~Gyr on the 
basis of H$\gamma_{HR}$. This difference could be attributed in part to the fact
that the stellar isochrones adopted in Worthey's (1994) population synthesis 
model are systematically hotter than those used in Vazdekis et al. (1996), and 
to possible effects caused by the fact that H$\gamma_{HR}$ is strongly sensitive
to very small $\sigma$ variations. However our age estimation is in full 
agreement with an extraordinary fit that we have obtained for this galaxy 
using V99 models (see Fig.~4 and Appendix).

The two clusters are found to be very old ($\stackrel{>}{_\sim}$13~Gyr). This 
sample of low-luminosity ellipticals show a rather large spread in age; 
NGC~4489 ($\sim3$~Gyr), M~32 ($\sim 4$~Gyr), NGC~4239 ($\sim 5.5$~Gyr), 
NGC~3605 ($\sim7.5$~Gyr) and NGC~4387 ($\sim12$~Gyr), respectively. Finally,
for NGC~7332 we obtain $\sim$6~Gyr (in full agreement with Vazdekis 1996). 
Since our results are based on SSP model predictions this method estimates 
luminosity-weighted average stellar ages, but cannot tell whether this spread 
is due to a real difference of {\it mean stellar age}, implying different 
epoch of galaxy formation, or caused by a contamination of intermediate age 
stars formed in a secondary episode of star formation (Kodama \& Arimoto 1998). 

[H$\gamma$+1/2(Fe{\sc I}+Mg{\sc I})]$_{\sigma}$ indices cover most 
of galaxy $\sigma$'s range (up to $300$~km$^{-1}$). In this paper we derive 
average ages for globular clusters and low-luminosity ellipticals. However we 
stress that these indices are fully applicable to normal and giant ellipticals, 
including cD's. Once high S/N spectra of relatively high 
dispersion are obtained, accurate and reliable estimates for the average 
stellar ages of these galaxies will be possible, without assessing their 
metallicities. Once the age is determined, metallicity can be evaluated 
uniquely from key metallic lines. The age-metallicity degeneracy of line 
indices is solved. 
We note that [H$\gamma$+1/2(Fe{\sc I}+Mg{\sc I})]$_{\sigma}$ indices are 
probably not unique as age indicators for old stellar
populations. Other Balmer lines with neighboring metallic lines could also
produce a similar effect and need to be addressed.

Direct estimates of galaxy age would provide a definitive answer to the origin 
of the CMR which is tightly followed by cluster elliptical galaxies (e.g., 
Bower, Lucey \& Ellis 1992). Field ellipticals and small groups 
of galaxies might have formed differently from those in rich clusters. 
Gonz\'alez (1993) showed that field ellipticals have significantly
stronger H$_{\beta}$ indices than expected from Worthey's (1994) models of 
very old age. This implies that they could be younger than 
cluster ellipticals. However H$_{\beta}$ is not an accurate age indicator 
since its dependence on metallicity is not negligible, and since it could be 
easily filled in with nebular emission (Davies et al. 1993) even 
for early-type galaxies. H$\gamma$ is substantially less affected. 

It is well known that stars in the Galactic halo, including those in globular 
clusters, are enhanced in $\alpha$-elements (e.g., Mg, Si, Ca) with respect 
to Fe (Edvardsson et al. 1993). Giant elliptical galaxies show 
$[\alpha/Fe] \simeq +0.3$ (Peletier 1989, Worthey, Faber \& Gonz\'alez 1992,
Vazdekis et al. 1997). This may cause a 
small effect on the resulting age estimate, because we have assumed solar 
abundance ratios throughout the present study (although 
[H$\gamma$+1/2(Fe{\sc I}+Mg{\sc I})]$_{\sigma}$ includes both Fe and
Mg in its definition). Worthey (1998) suggested that abundance ratio changes
can cause changes in isochrone temperature structure that can mimic an age 
effect. Salaris \& Weiss (1998) have shown that scaled-solar isochrones no longer
can be used to replace $\alpha$-enhanced ones at the same total metallicity.
Unfortunately, empirical spectra of stars with different $[\alpha/Fe]$ ratios 
are not yet available and we cannot assess this problem for the moment.

\section{Conclusions}
We define new spectral indices, [H$\gamma$+1/2(Fe{\sc I}+Mg{\sc I})]$_{\sigma}$, 
which are robust age indicators for old and relatively old stellar populations 
and therefore have a great potential for breaking the age-metallicity degeneracy 
of galaxy spectra. To achieve this we use the new evolutionary stellar 
population synthesis model of Vazdekis (1999) which provides SED's of SSP's at 
FWHM$\sim1.8$\AA. These new models not only allow us to investigate the behavior
of H$\gamma$ as a function of metallicity and age, but also as a function of 
spectral resolution by performing a direct measurement of the feature on the 
synthesized SSP spectra broadened to different $\sigma$'s. We show that the 
strong power of H$\gamma$ to disentangle the age is due to a compensating 
effect: at specified age the strengthening of H$\gamma$ is mainly caused by 
its adjacent metallic absorption which is more prominent for more metal-rich 
populations. On the other hand the adopted pseudocontinua are depressed by the 
effects of the strong neighboring Fe{\sc I} lines on both sides of H$\gamma$ 
(centered on 4325\AA~and 4352\AA). On the basis of this effect we achieve 
new H$\gamma$ index definitions that are completely insensitive to the metallicity 
in the range -0.7$\leq$[Fe/H]$\leq$+0.2. However, since this effect is strongly 
dependent on the adopted resolution and galaxy velocity dispersion,
we have optimized H$\gamma$ definitions for assessing the resolution stability 
required for practical applications. We propose unprecedented age 
indicators, [H$\gamma$+1/2(Fe{\sc I}+Mg{\sc I})]$_{\sigma}$, at well separated 
velocity dispersions: $\sigma=125$, $200$, and $275$~kms$^{-1}$, and each is 
stable in a range of $\Delta\sigma\sim75$~kms$^{-1}$, thus allowing study of 
most galaxies up to $\sigma\sim300$~kms$^{-1}$. 

The main characteristics and uncertainties affecting
[H$\gamma$+1/2(Fe{\sc I}+Mg{\sc I})]$_{\sigma}$ are extensively discussed. Since 
we are dealing with the absorption profile of H$\gamma$, observational spectra of 
very high signal-to-noise (S/N(\AA)$\stackrel{>}{_\sim}$200) and relatively high 
dispersion ($\stackrel{<}{_\sim}$0.8\AA/pix), but feasible for luminous nearby 
galaxies with 4m class telescopes, are required to gain this unprecedented age 
discriminating power. We provide a recipe for a correct measurement of
[H$\gamma$+1/2(Fe{\sc I}+Mg{\sc I})]$_{\sigma}$: determine $\sigma$ of the 
observed spectrum, determine its absolute radial velocity (by using a synthetic 
spectrum which match best the age/metallicity of the galaxy), measure the index 
appropriate for that velocity dispersion, and compare with the
metallicity-insensitive model values of Table~2 to estimate the age. High
S/N's can be achieved by integrating along the spatial direction if variations of 
velocity rotation (and $\sigma$) are properly accounted for.

Once observational spectra of high quality are acquired, accurate and reliable 
estimates for the luminosity-weighted average stellar ages of these galaxies 
will be possible for the first time, without assessing their metallicities. 
In this paper we perform such measurements for two metal-rich globular clusters,
a sample of low-luminosity ellipticals and a standard S0 galaxy. We find a 
large spread in the average stellar ages of these low-luminosity ellipticals. 
In particular, these indices yield 4~Gyr for M~32. This value is in excellent 
agreement with the age provided by an extraordinary fit to the full spectrum 
of this galaxy that we achieve in this paper.

\acknowledgments

We are indebted to J. Rose and L. Jones for providing their globular
cluster and elliptical galaxy spectra. We are grateful to V.Vansevi\v cius 
and R. Peletier for very interesting comments and suggestions. We are also 
grateful to the referee for useful corrections and suggestions. A.V. thanks 
the Japan Society for Promotion of Science for financial support. This work 
was financially supported in part by a Grant-in-Aid for the Scientific Research
(No.09640311) by the Japan Ministry of Education, Culture, Sports and Science.

\appendix
\section{Accurate absolute radial velocity determination for measuring 
[H$\gamma$+1/2(Fe{\sc I}+Mg{\sc I})]$_{\sigma}$}
Accurate radial velocity determinations can be achieved if an observational 
galaxy spectrum is crosscorrelated with a template which resembles best the
galaxy (e.g. Tonry \& Davis 1979). This approach would benefit substantially if
templates are chosen from the new SSP model spectral library of V99, rather 
than using observations of a template star (which could easily be a poor match, 
thus producing a crosscorrelation peak with asymmetrical wings that could bias 
the centroid of the fitting function, e.g., gaussian). Therefore we strongly 
suggest the use as templates of at least various synthetic spectra
(FWHM$\sim$1.8\AA) of different ages (e.g., 1.5, 3, 6 and 12~Gyr) and 
metallicities ([Fe/H]=-0.7, -0.4, 0.0, +0.2) which can be retrieved from our
web site (http://www.ioa.s.u-tokyo.ac.jp/$\sim$vazdekis/). Choosing the model
spectrum which provides the largest crosscorrelation peak height (i.e., 
produced by the most appropriate template), we determine accurately 
the absolute radial velocity and prevent any additional shifts to the models.
Prior to crosscorrelation all the spectra should be rebinned logarithmically (with
same dispersion), continuum removed (e.g., using a spline3 of low order).
Results improve if the spectra are adequately filtered to remove their very
high and very low frecuencies, and multiplied by a cosine-bell-like function.
Detailed explanations can be found in V99 (and references therein).

As an example, we have crosscorrelated M~32 spectrum (Jones 1997) with the 
whole SSP model spectral library of V99. Fig.~4 shows that the largest 
crosscorrelation peak height is obtained for a model of $\sim4$~Gyr and
[Fe/H]=0.0. This implies that the use of this spectrum (or a similar one) as 
template would increase the accuracy of the kinematical parameter determination.
This model is overplotted on M~32 spectrum providing an extraordinary 
fit (see Fig.~4) which confirms the age predicted by 
[H$\gamma$+1/2(Fe{\sc I}+Mg{\sc I})]$_{\sigma}$.

\newpage
\figcaption{H${\gamma}$$_{HR}$ index measured on V99 model spectral library 
broadened to various resolutions: 
$\sigma=60$~kms$^{-1}$, $125$~kms$^{-1}$, $200$~kms$^{-1}$, and $275$~kms$^{-1}$}

\figcaption{Model spectra of 13~Gyr and [Fe/H]$=-0.7$ (dot-dashed lines), 
[Fe/H]$=-0.4$ (dotted lines), [Fe/H]$=0.0$ (solid lines) and [Fe/H]$=+0.2$ 
(dashed lines), broadened to the resolutions used in Fig~1. All these spectra 
were normalized to 1 at 4365\AA. On the first panel we mark H${\gamma}_{HR}$ 
definition of JW95 (the arrows indicate the positions of the adopted continua), 
while on the other panels we mark 
[H$\gamma$+1/2(Fe{\sc I}+Mg{\sc I})]$_{\sigma}$ definition for each resolution: 
thick dotted lines represent the feature while thick solid lines represent the 
pseudocontinua}

\figcaption{[H$\gamma$+1/2(Fe{\sc I}+Mg{\sc I})]$_{\sigma}$ indices measured on 
the same SSP's spectra of Fig.~1. On the first panel, open stars represent 
[H$\gamma$+1/2(Fe{\sc I}+Mg{\sc I})]$_{125}$ values measured on the spectra of 
two metal-rich globular clusters (47~Tuc and NGC~6624). Open circles are the 
values obtained for a sample of low-luminosity ellipticals of Jones (1997). 
Open polygon represents the value for the central region of the NGC~7332 S0 
galaxy (Vazdekis 1996). The largest error bar is also plotted. Table~2 was used 
to infer the age. M~32 spectrum was broadened to $\sigma\sim200$~kms$^{-1}$ and 
$\sim275$~kms$^{-1}$ for measuring [H$\gamma$+1/2(Fe{\sc I}+Mg{\sc I})]$_{200}$ 
(upper-right panel) and [H$\gamma$+1/2(Fe{\sc I}+Mg{\sc I})]$_{275}$ 
(bottom-right panel) respectively, providing the same age (4~Gyr). Finally, the 
bottom-left panel represents a test to the stability of 
[H$\gamma$+1/2(Fe{\sc I}+Mg{\sc I})]$_{125}$; measured on the model spectra 
after broadening to $\sigma=100$~kms$^{-1}$ (thin lines for different 
metallicities), $\sigma=125$~kms$^{-1}$ (thick lines, i.e. the same values as 
in the upper-left panel) and $\sigma=175$~kms$^{-1}$ (very thick lines)}

\figcaption{Upper panel shows the crosscorrelation peak height of M~32 spectrum 
with V99 model spectral library. All these spectra were broadened to 
$\sigma=125$~kms$^{-1}$. Lower panel shows M~32 spectrum and the synthetic spectrum 
which provided the largest crosscorrelation peak height, achieving an extraordinary 
fit}

\newpage
\begin{table}
\begin{center}
\begin{tabular}{lccc}
\hline\hline
Index &Blue pseudocont. &Feature &Red pseudocont.\\
\hline
$[H\gamma$+1/2(Fe{\sc I}+Mg{\sc I})$]_{125}$ &4330.000 4340.468 &4333.000 
4352.737 &4359.250
4368.750\\
$[H\gamma$+1/2(Fe{\sc I}+Mg{\sc I})$]_{200}$ &4331.000 4340.750 &4332.000 
4352.250 &4359.250
4368.750\\
$[H\gamma$+1/2(Fe{\sc I}+Mg{\sc I})$]_{275}$ &4331.500 4341.000 &4331.500 
4351.875 &4359.250
4368.750\\
\end{tabular}
\end{center}
\caption{[H$\gamma$+1/2(Fe{\sc I}+Mg{\sc I})]$_{\sigma}$ index definitions}
\end{table}

\begin{table}
\begin{center}
\begin{tabular}{rccc}
\hline\hline
{\tiny Age}&{\tiny [H$\gamma$+1/2(Fe{\sc I}+Mg{\sc I})]$_{125}$}&{\tiny
[H$\gamma$+1/2(Fe{\sc I}+Mg{\sc
I})]$_{200}$}&{\tiny [H$\gamma$+1/2(Fe{\sc I}+Mg{\sc I})]$_{275}$}\\
\hline
 1.6& 1.366& 0.824& 0.534\\
 2.0& 1.289& 0.773& 0.498\\
 2.5& 1.186& 0.702& 0.446\\
 3.0& 1.140& 0.672& 0.425\\
 4.0& 1.062& 0.619& 0.387\\
 5.0& 1.021& 0.592& 0.368\\
 6.0& 0.997& 0.575& 0.357\\
 7.0& 0.965& 0.553& 0.342\\
 8.0& 0.929& 0.530& 0.325\\
 9.0& 0.908& 0.516& 0.316\\
10.0& 0.887& 0.502& 0.306\\
11.0& 0.869& 0.492& 0.299\\
12.0& 0.851& 0.480& 0.291\\
13.0& 0.843& 0.476& 0.288\\
14.0& 0.833& 0.469& 0.283\\
15.0& 0.828& 0.466& 0.280\\
16.0& 0.820& 0.461& 0.278\\
17.0& 0.812& 0.457& 0.274\\
17.4& 0.809& 0.455& 0.273\\
\end{tabular}
\end{center}
\caption{Predicted index values for different ages. The tabulated numbers are 
obtained by averaging the index values obtained for different metallicities.
The $\sigma$ ranges in which these indices are very stable against resolution
are: $115-160$~kms$^{-1}$ for 
[H$\gamma$+1/2(Fe{\sc I}+Mg{\sc I})]$_{125}$, $165-210$~kms$^{-1}$ for 
[H$\gamma$+1/2(Fe{\sc I}+Mg{\sc I})]$_{200}$ and $240-285$~kms$^{-1}$ for 
[H$\gamma$+1/2(Fe{\sc I}+Mg{\sc I})]$_{275}$}
\end{table}

\begin{table}
\begin{center}
\begin{tabular}{lccc}
\hline\hline
&{\tiny [H$\gamma$+1/2(Fe{\sc I}+Mg{\sc I})]$_{125}$}&{\tiny 
[H$\gamma$+1/2(Fe{\sc
I}+Mg{\sc
I})]$_{200}$}&{\tiny [H$\gamma$+1/2(Fe{\sc I}+Mg{\sc I})]$_{275}$}\\
\hline
Optimal spectral resolution $\sigma$(kms$^{-1}$) &125       &200       &275    \\
Stability $\sigma$(kms$^{-1}$) range             &100-175   &150-225   &225-300\\
Recommended $\sigma$(kms$^{-1}$) range           &115-160   &165-210   &240-285\\
Required S/N per \AA                           &$\sim$200&$\sim$300&$\sim$400\\
Exp.time($\mu_{B}$=17.5,Int.=2$\arcsec$,Tel.=4.2m,0.8\AA/pix,$\sigma_{RN}$=4.4)&
$\sim$1.5hour&$\sim$3hour&$\sim$5hour\\
Maximum $\Delta\lambda$ shift (e.g., errors in $\lambda$, $z$, rot. curve)&0.12\AA &0.08\AA&0.07\AA\\
$\Delta$[H$\gamma$+1/2(Fe{\sc I}+Mg{\sc I})]$_{\sigma}$ (Flux calib. - Continuum
remov.)&$<$0.022\AA&$<$0.015\AA&$<$0.009\AA\\
Largest theoretical age uncertainty     &14-17.5Gyr &14-17.5Gyr &14-17.5Gyr\\
Index error causing an age uncertainty of 11.5-17.5Gyr &$\sim$0.030\AA &$\sim$0.021\AA &$\sim$0.017\AA\\
\end{tabular}
\end{center}
\caption{Index characteristics and uncertainties}
\end{table}

\addtocounter{figure}{-4}

\newpage
\begin{figure}
\plotone{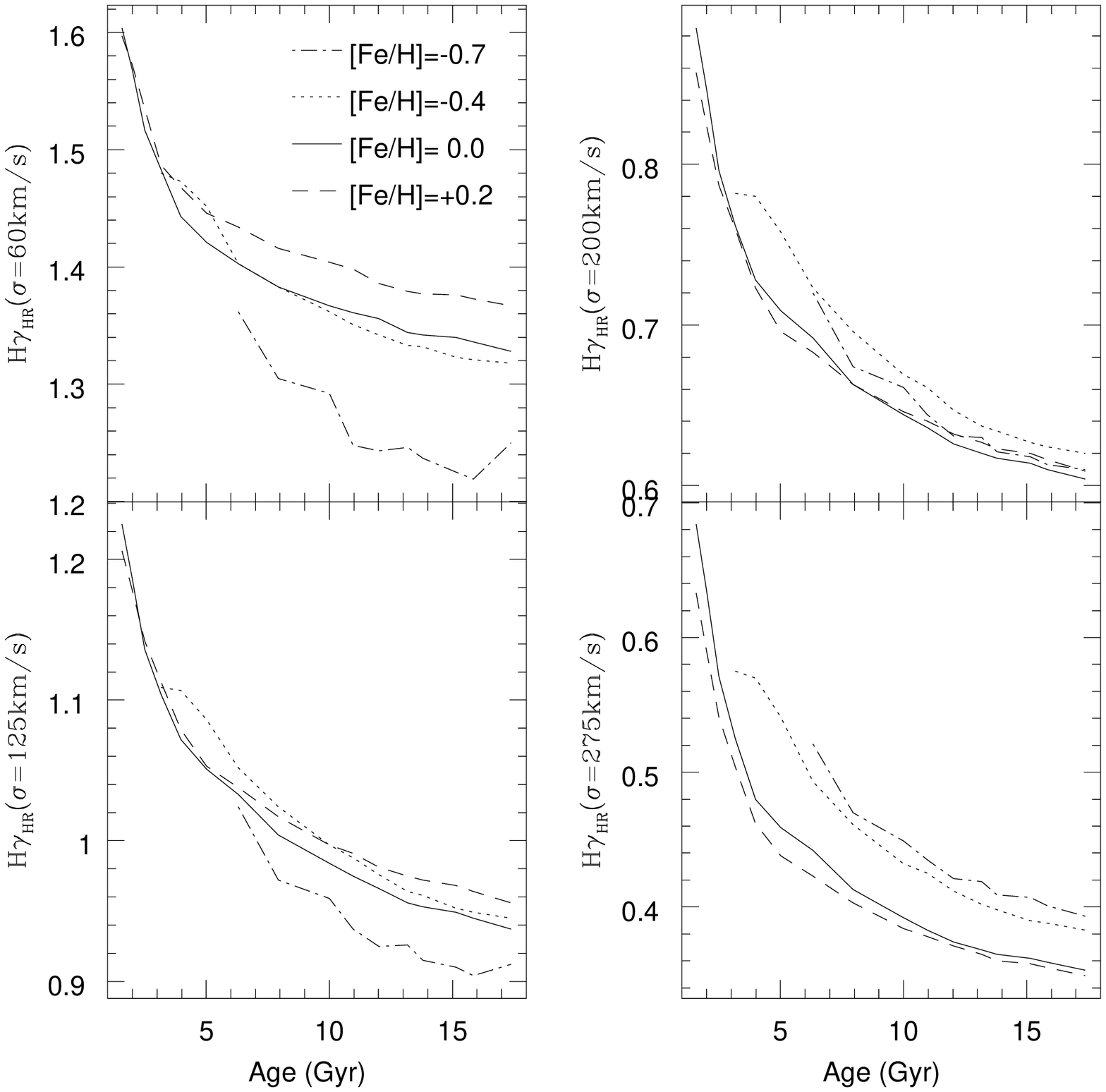}
\caption{}
\end{figure}

\begin{figure}
\plotone{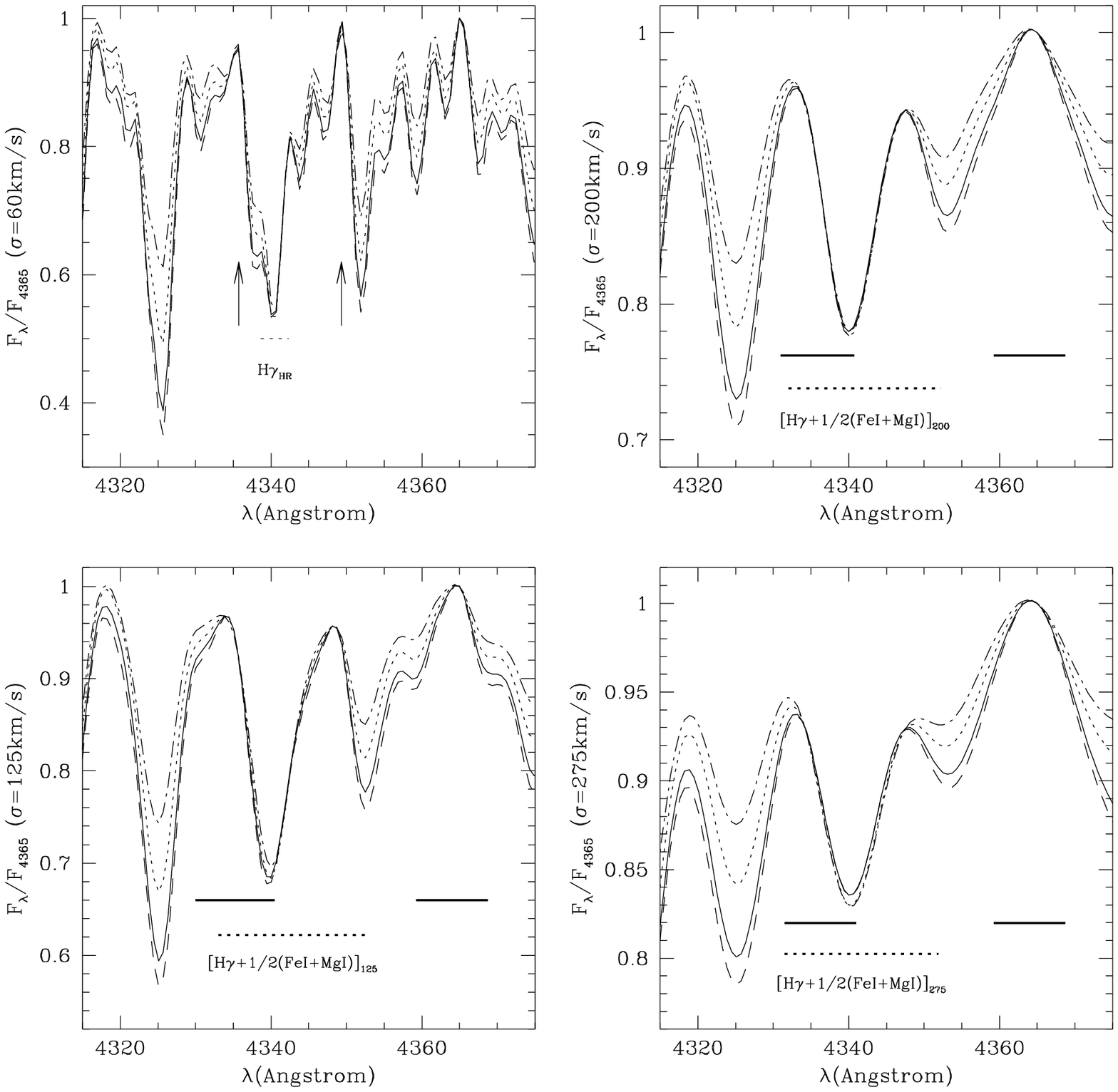}
\caption{}
\end{figure}

\begin{figure}
\plotone{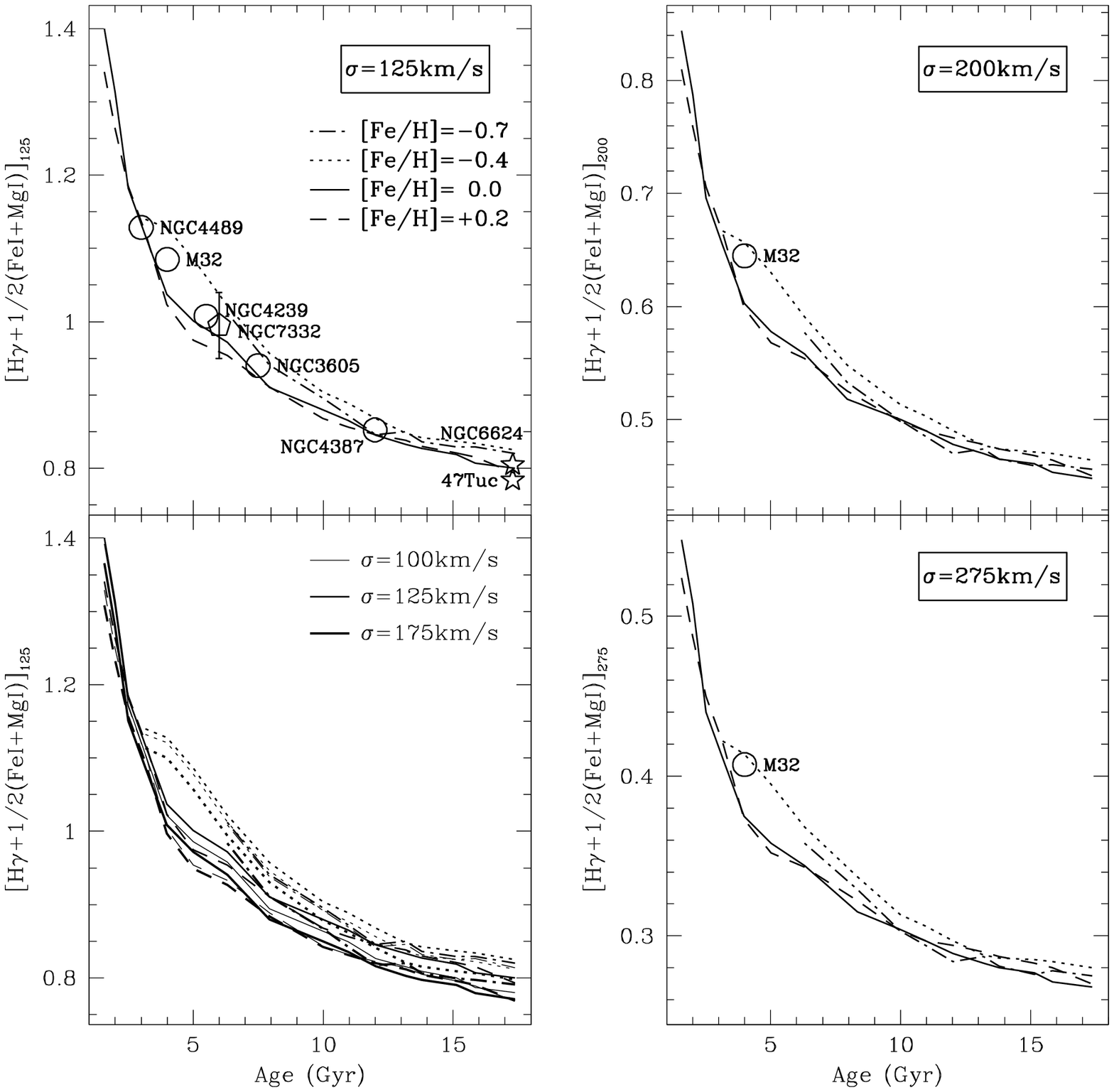}
\caption{}
\end{figure}

\begin{figure}
\plotone{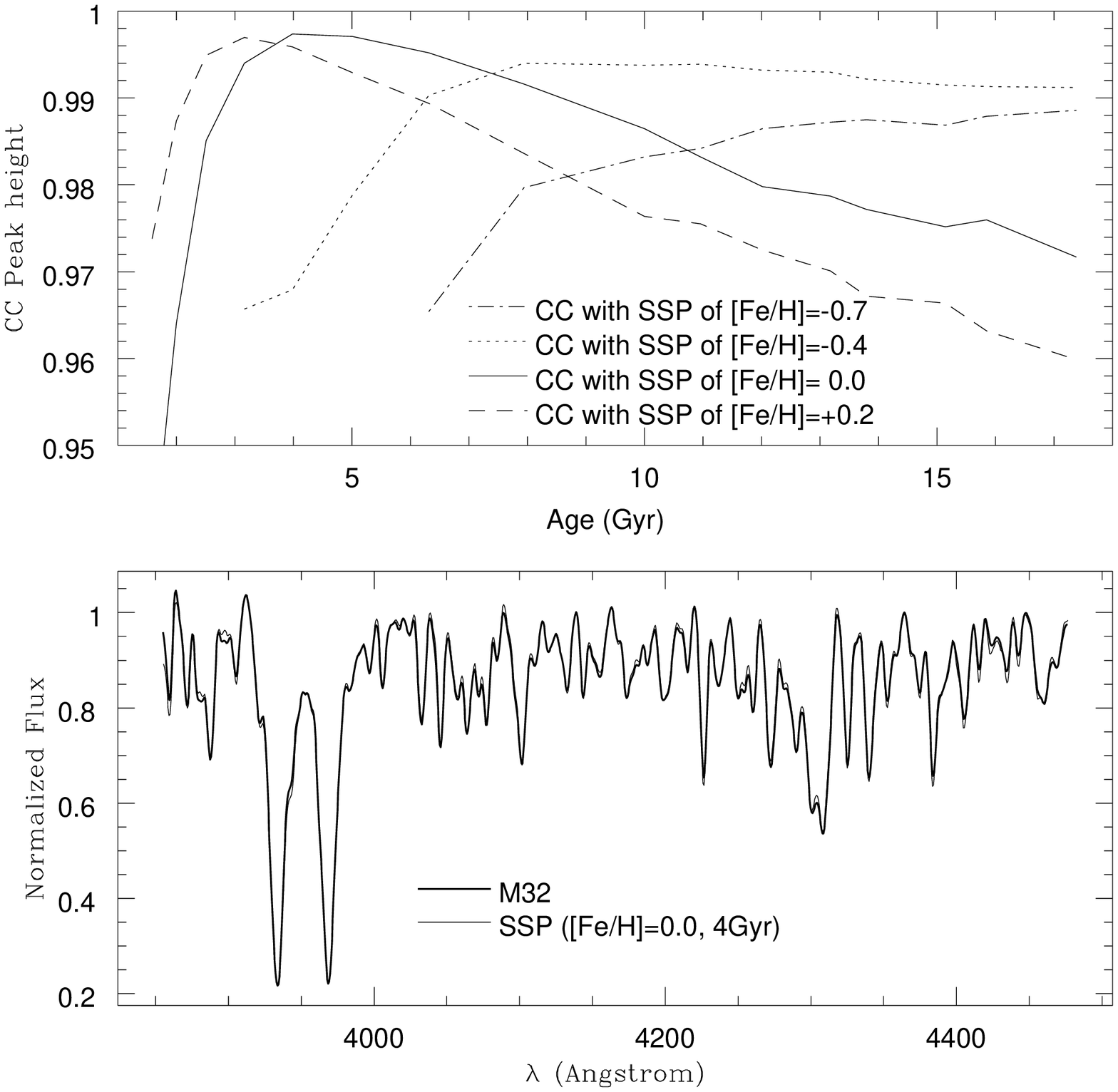}
\caption{}
\end{figure}

\end{document}